\documentclass[aps,prd,showpacs,twocolumn,groupedaddress,nofootinbib]{revtex4}
\usepackage{graphicx}
\usepackage{amsmath}
\usepackage{epsfig}
\usepackage{psfig}

{
{

\begin{document}

\title{A Minimal Type II Seesaw Model}

\author{Pei-Hong Gu}
\email{guph@mail.ihep.ac.cn}

\author{He Zhang}
\email{zhanghe@mail.ihep.ac.cn}

\author{Shun Zhou}
\email{zhoush@mail.ihep.ac.cn}

\affiliation{Institute of High Energy Physics, Chinese Academy of
Sciences, P.O. Box 918 (4), Beijing 100049, China}
\thanks{Mailing address\\}

\begin{abstract}
We propose a minimal type II seesaw model by introducing only one
right-handed neutrino besides the $SU(2)_{L}$ triplet Higgs to the
standard model. In the usual type II seesaw models with several
right-handed neutrinos, the contributions of the right-handed
neutrinos and the triplet Higgs to the CP asymmetry, which stems
from the decay of the lightest right-handed neutrino, are
proportional to their respective contributions to the light
neutrino mass matrix. However, in our minimal type II seesaw
model, this CP asymmetry is just given by the one-loop vertex
correction involving the triplet Higgs, even though the
contribution of the triplet Higgs does not dominate the light
neutrino masses. For illustration, the Fritzsch-type lepton mass
matrices are considered.

\end{abstract}


\maketitle

Recent neutrino oscillation experiments have provided us with very
convincing evidence that neutrinos are massive and lepton flavors
are mixed \cite{sv2006}. In the three-neutrino mixing scheme, a
global analysis of experimental data yields: two independent
neutrino mass-squared differences $\Delta m^2_{21} = (7.2 \sim
8.9) \times 10^{-5} ~ {\rm eV}^2$, $|\Delta m^2_{32}| = (1.7 \sim
3.3) \times 10^{-3} ~ {\rm eV}^2 $ and three mixing angles
$30^\circ < \theta^{}_{12} < 38^\circ$, $36^\circ < \theta^{}_{23}
< 54^\circ$, $\theta^{}_{13} < 10^\circ$ at the $99\%$ confidence
level \cite{Data}.

On the other hand, the cosmological baryon asymmetry, which is
characterized by the ratio of baryon to photon number densities,
has also been measured by the WMAP experiment to a very good
precision \cite{WMAP}
\begin{equation}
\label{eta} \eta^{}_{\rm B} \equiv \frac{n^{}_{\rm
B}}{n^{}_{\gamma}} = (6.1 \pm 0.2)\times 10^{-10} \; .
\end{equation}
Sometimes the baryon asymmetry is also represented by $Y^{}_{\rm
B}\equiv n^{}_{\rm B}/s = \eta^{}_{\rm B}/7.04 \simeq (8.4 \sim
8.9) \times 10^{-11}$ with $s$ being the entropy density.

Leptogenesis \cite{leptogenesis} is now an attractive scenario to
simultaneously explain the neutrino oscillation phenomena and the
baryon asymmetry through the famous seesaw mechanism \cite{SS}. In
the ordinary type I seesaw models, the leptogenesis scenario can
be realized by introducing two or more right-handed neutrinos with
heavy Majorana masses to the $SU(2)_{L}\times U(1)_{Y}$ standard
model. Another interesting alternative is to consider the
so-called type II seesaw models \cite{typeII}, in which the
$SU(2)_{L}$ triplet Higgs, besides the right-handed neutrinos, can
also contribute to the light neutrino masses as well as the baryon
asymmetry generation.

In this paper, we propose a minimal type II seesaw model by
extending the standard model with only one right-handed neutrino
in addition to the $SU(2)_{L}$ triplet Higgs.\footnote{Here the
{\it minimal} is in the sense of the number of new particles
compared with the conventional type II seesaw model.} The CP
asymmetry in the decay of the right-handed neutrino just arises
from the interference between the tree level diagram and the
one-loop vertex correction involving the triplet Higgs. At the
same time, the light neutrino mass matrix can receive comparable
contributions from the triplet Higgs and the right-handed
neutrino. For illustration, we will consider Fritzsch-type lepton
mass matrices with five texture zeroes in the specific discussions
and calculations. It will be shown that our model can
simultaneously explain the neutrino properties and the baryon
asymmetry.

It is convenient to start with our discussions from the general type
II seesaw models \cite{Kang,hs2004}:
\begin{eqnarray}
\label{lagrangian} -\mathcal{L} &=&
M_{\Delta}^{2}\textrm{Tr}\Delta_{L}^{\dagger}\Delta_{L}^{}
+g_{\alpha\beta}\bar{\psi}^{C}_{L\alpha}i\tau_{2}\Delta_{L}^{}\psi_{L\beta}^{}
-\mu\phi^{T}i\tau_{2}\Delta_{L}\phi^{} \nonumber \\
&&+ \frac{1}{2}M_{i}\bar{\nu}^{C}_{Ri}\nu_{Ri}^{} +y_{\alpha
i}\bar{\psi}_{L\alpha}\nu_{Ri}^{}\phi^{} +h.c.\, \nonumber\\
&=&
M_{\Delta}^{2}\textrm{Tr}\Delta_{L}^{\dagger}\Delta_{L}^{}+\frac{1}{2}M_{i}\bar{N}_{i}N_{i}
+g_{\alpha\beta}\bar{\psi}^{C}_{L\alpha}i\tau_{2}\Delta_{L}^{}\psi_{L\beta}^{} \nonumber \\
&&- \mu\phi^{T}i\tau_{2}\Delta_{L}^{}\phi^{}+ y_{\alpha
i}\bar{\psi}_{L\alpha}N_{i}\phi^{} + {\rm h.c.}\,.
\end{eqnarray}
Here $\psi_{\alpha}=(\nu_{\alpha}, l_{\alpha})^{T}\,
(\alpha=e,\mu,\tau)$, $\phi=(\phi^{0}, \phi^{-})^{T}$ are the
lepton and the Higgs doublets, while
\begin{displaymath}
\label{delta}
\Delta_{L}=  \left(\begin{array}{cc}
\frac{1}{\sqrt{2}}\delta^{+} & \delta^{++} \\
\delta^{0} & -\frac{1}{\sqrt{2}}\delta^{+}
\end{array}\right)
\end{displaymath}
is the triplet Higgs. $\nu_{Ri}^{}\, (i=1,...,d)$ are the
right-handed neutrinos, and $N_{i}=\nu_{Ri}^{}+ \nu_{Ri}^{C}$ is
defined as the heavy Majorana neutrinos. We have conveniently
chosen the basis, in which the Majorana mass matrix of
right-handed neutrinos is diagonal, \textit{i.e.}
$M=\textrm{Diag}\{ M_{1},...,M_{d}\}$. Obviously, $g$ is generally
a complex $3\times 3$ matrix. In the usual type II seesaw models,
there are several right-handed neutrinos, \textit{i.e.} $d\geq 2$.
However, only one right-handed neutrino and hence three Yukawa
couplings $y^{}_{\alpha 1}\, (\alpha=e,\mu,\tau)$ occur in our
minimal type II seesaw model with $d=1$.

After the electroweak phase transition, the mass matrix of the
light neutrinos $\nu_{\alpha}=\nu_{L\alpha}^{}+ \nu_{L\alpha}^{C}$
can be written as
\begin{equation}
\label{lightmass}
M_{\nu}=-y^{*}\frac{1}{M}y^{\dagger}v^{2}+2gv_{L}=M_{\nu}^{\textrm{I}}+M_{\nu}^{\textrm{II}}\,,
\end{equation}
where $M_{\nu}^{\textrm{I}}$ is the ordinary type I seesaw mass
term, $M_{\nu}^{\textrm{II}}$ is the type II seesaw mass term.
$v=174\,\textrm{GeV}$, $v_{L}\simeq\mu^{*}v^{2}/M_{\Delta}^{2}$
are the vacuum expectation values of $\phi$ and $\Delta_{L}^{}$,
respectively. It is easy to see that $v_{L}$ is naturally seesaw
suppressed if $\Delta_{L}$ is very heavy.

\begin{figure}\vspace{3.0cm}
\psfig{file=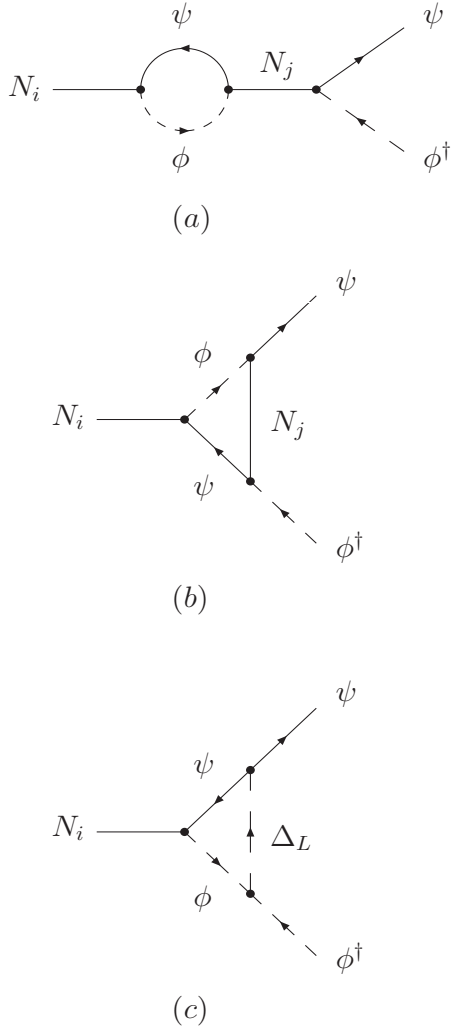, bbllx=7.5cm, bblly=3.0cm, bburx=22cm, bbury=21.0cm,%
width=12cm, height=14cm, angle=0, clip=0} \vspace{-2.0cm}
\caption{\label{loop}
 The one-loop diagrams of $N_{i}$ decay.}
\end{figure}

The CP asymmetry $\varepsilon_{N_{i}}^{}$ from the decay of the
heavy Majorana neutrinos $N_{i}$ is given by the interference of the
ordinary tree level decay with the three diagrams of Fig.
\ref{loop}. The first two diagrams are the self-energy and vertex
correction involving the heavy Majorana neutrinos, while the third
is due to the contribution from the triplet Higgs. Under the
assumption that $M_{1}\ll M_{2},...,M_{d}, M_{\Delta}$, the CP
asymmetry $\varepsilon_{N_{1}}$ can be simplified as
\cite{hs2004,ak2004}:
\begin{eqnarray}
\label{totalcp} \varepsilon_{N_{1}}^{} &=&
\varepsilon_{N_{1}}^{N}+\varepsilon_{N_{1}}^{\Delta}\,,\\
\label{cpN} \varepsilon_{N_{1}}^{N}& \simeq & \frac{3}{16
\pi}\frac{M_{1}}{v^{2}}\frac
{\sum_{\alpha\beta}\text{Im}\left[y^{\nu\dagger}_{1\alpha}y^{\nu\dagger}_{1\beta}\left(M^{\textrm{I}*}_{\nu
}\right)_{\alpha\beta}\right]}{\left(y^{\nu\dagger}y^{\nu}\right)_{11}}\,,\\
\label{cpDelta} \varepsilon_{N_{1}}^{\Delta} & \simeq &
\frac{3}{16 \pi} \frac{M_{1}}{v^{2}} \frac{\sum_{\alpha\beta}
\text{Im}\left[y^{\nu \dagger}_{1\alpha}y^{\nu \dagger}_{1\beta}
\left(M^{\textrm{II}*}_{\nu}\right)_{\alpha\beta}\right]}{\left(y^{\nu
\dagger}y^{\nu}\right)_{11}}\,.
\end{eqnarray}
It is shown that the contributions of $N_{i}$ (diagrams (a) and
(b) in Fig. \ref{loop}) and $\Delta_{L}$ (diagram (c) in Fig.
\ref{loop}) to $\varepsilon_{N_{1}}$ are proportional to their
respective contributions to the light neutrino mass matrix
\cite{hs2004}. Therefore, the contribution from the triplet Higgs
to the above CP asymmetry can be neglected in the limit
$M_{\nu}^{\textrm{II}}\ll M_{\nu}^{\textrm{I}}$. However, this
analysis is invalid in our minimal type II seesaw model, since
there is just one heavy Majorana neutrino, the one-loop vertex
correction involving the triplet Higgs is the sole source of this
CP asymmetry and we always obtain
\begin{eqnarray}
\label{cpMinimal} \varepsilon_{N_{1}}^{} \equiv
\varepsilon_{N_{1}}^{\Delta}\,,
\end{eqnarray}
even though the contribution of $\Delta_{L}$ does not dominate the
light neutrino mass.

\begin{figure}\vspace{-0.5cm}
\psfig{file=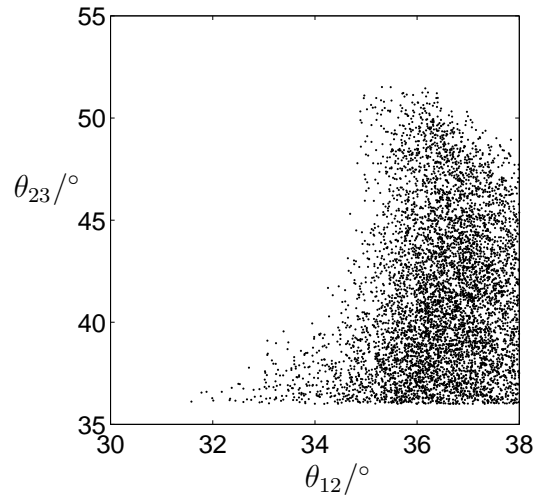, bbllx=4.0cm, bblly=5.0cm, bburx=14cm, bbury=16.0cm,%
width=10cm, height=11cm, angle=0, clip=0} \vspace{-3.5cm}
\caption{\label{theta12} Allowed region of solar and atmospheric
mixing angles, namely $\theta^{}_{12}$ and $\theta^{}_{23}$ in the
standard parametrization \cite{PDG}. }
\end{figure}

\begin{figure}
\psfig{file=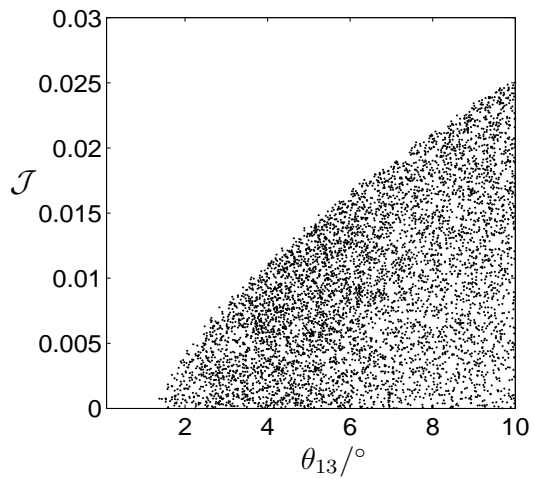, bbllx=4.0cm, bblly=5.0cm, bburx=14cm, bbury=16.5cm,%
width=10cm, height=11cm, angle=0, clip=0} \vspace{-3.5cm}
\caption{\label{theta13} Allowed region of the mixing angle
$\theta^{}_{13}$ and Jarlskog invariant $\cal J$. }
\end{figure}

For illustration, we simply assume the Fritzsch-type textures
\cite{Fritzsch} of $M^{\rm II}_{\nu}$ and the charged lepton mass
matrix $M^{}_l$:
\begin{equation}
\label{massII} M^{\rm II}_\nu = v^{}_{L} \left(
\begin{matrix} 0 & C^{}_\nu e^{i \alpha^{}_{\nu}} & 0 \cr C^{}_\nu
e^{i \alpha^{}_{\nu}} & 0 & B^{}_\nu e^{i\beta^{}_{\nu}} \cr 0 &
B^{}_\nu e^{i \beta^{}_{\nu}} & A^{}_\nu e^{i
\gamma^{}_{\nu}}\end{matrix} \right)
\end{equation}
and
\begin{equation}
\label{massLepton} M_{l} =  v \left( \begin{matrix} 0 & C^{}_l
e^{i \alpha^{}_l} & 0 \cr C^{}_l e^{i \alpha^{}_l} & 0 & B^{}_l
e^{i\beta^{}_l} \cr 0 & B^{}_l e^{i \beta^{}_l} & A^{}_l e^{i
\gamma^{}_l} \end{matrix} \right) \ ,
\end{equation}
where $A^{}_{\nu,l}$, $B^{}_{\nu,l}$ and $C^{}_{\nu,l}$ are real
and positive. In addition, we set
\begin{eqnarray}
\label{yukawa} y = i y^{}_{0} \left(0, r , 1 \right)^T \,,
\end{eqnarray}
where the imaginary unit $i$ has been inserted to cancel the minus
sign in front of the type I term for convenience. Substituting
Eqs. (\ref{massII}) and (\ref{yukawa}) into Eq. (\ref{lightmass}),
we can obtain the effective neutrino mass matrix
\begin{equation}
\label{massNeutrino} M^{}_\nu = m_0 \left(\begin{matrix} 0 &
\widehat {C} e^{i \alpha_\nu} & 0 \cr \widehat {C} e^{i
\alpha_\nu} & r^2  & r +\widehat {B} e^{i \beta_\nu}  \cr 0 & r
+\widehat {B} e^{i \beta_\nu} & 1 + \widehat {A} e^{i \gamma_\nu}
\end{matrix} \right) \ ,
\end{equation}
where $m^{}_0 \equiv v^2 y_{0}^2/M^{}_{1}$ and $\widehat{A} \equiv
v^{}_{L} A^{}_\nu /m^{}_0$, likewise for $\widehat{B}$ and
$\widehat{C}$. The strategies to diagonalize lepton mass matrices
of the form in Eqs. (\ref{massLepton}) and (\ref{massNeutrino})
can be found in Refs. \cite{Fritzsch} and \cite{Zhang}. For
simplicity, we adopt two assumptions: $r = \sqrt{m^{}_2/m^{}_0}$
and $\arg \left( 1 + \widehat {A} e^{i \gamma_\nu} \right)=2 \arg
\left(  r  + \widehat {B} e^{i \beta_\nu} \right)$, and then
obtain
\begin{eqnarray}
\label{ABCnu} \widehat{A} & = & \left[\frac{(m_3-m_1)^2}{m^2_0}
-\sin^2{\gamma_\nu}\right]^{1/2}
-\cos{\gamma_\nu} \,, \nonumber \\
\widehat{B} & = & \left[ \frac{m_1 m_3 (m_3 -m_1
-m_2)}{m^2_0(m_3-m_1)} -r^2 \sin^2\beta_\nu
\right]^{1/2} \nonumber \\
&&-r \cos\beta_\nu \,, \nonumber \\
\widehat{C} & = & \left[\frac{m_1 m_2 m_3}{m^2_0
(m_3-m_1)}\right]^{1/2} \;
\end{eqnarray}
and
\begin{eqnarray}
\label{ABCl}
{A}_l & = & \left(m_\tau - m_\mu + m_e\right) \,, \nonumber \\
{B}_l & = & \left[\frac{(m_\mu-m_e)(m_\tau-m_\mu)(m_e+m_\tau)}{ (m_\tau-m_\mu+m_e)} \right]^{1/2}\,, \nonumber \\
{C}_{l} & = & \left[\frac{m_e m_\mu
m_\tau}{(m_\tau-m_\mu+m_e)}\right]^{1/2}\,.
\end{eqnarray}
The Maki-Nakagawa-Sakata (MNS) mixing matrix is $V = U^\dagger_l
U^{}_\nu$, where $U^\dagger_l$ and $ U^{}_\nu$ are the unitary
matrices used to diagonalize the lepton mass matrices:
$U^\dagger_l M^{}_l U^*_l = {\rm Diag}\left\{m^{}_e, m^{}_\mu,
m^{}_\tau \right\}$ and $U^\dagger_\nu M^{}_\nu U^*_\nu = {\rm
Diag}\left\{m^{}_1, m^{}_2, m^{}_3 \right\}$. Taking the neutrino
oscillation experimental data at the $99\%$ confidence level as
input\footnote{For simplicity, we assume the normal mass hierarchy
of light neutrinos in our numerical analysis. }, the allowed
regions of three mixing angles and Jarlskog invariant ${\cal J}
\equiv {\rm Im} \left(V^{}_{11} V^{}_{22} V^*_{12} V^*_{21}
\right) $ in our model are shown in Figs. \ref{theta12} and
\ref{theta13}. The phase parameters, exactly combinations of
$\alpha^{}_{l, \nu}, \beta^{}_{l, \nu}$ and $\gamma^{}_{l, \nu}$,
are all varying in the range $[0, 2\pi)$. From Fig. 3, we can see
that the smallest mixing angle $\theta^{}_{13}$ is larger than
$1^\circ$ and the maximal value of CP-violation parameter $\cal J$
can reach $2.5\%$, which are to be strictly tested in the future
neutrino oscillation experiments. Here we remark that the
Fritzsch-type lepton mass matrices can serve as a good example to
illustrate the features of the minimal type II seesaw model. For
more phenomenological discussions about lepton mass matrices in
the type-II seesaw models, see \cite{Guo}.


\begin{figure}[htbp]\vspace{-0.5cm}
\psfig{file=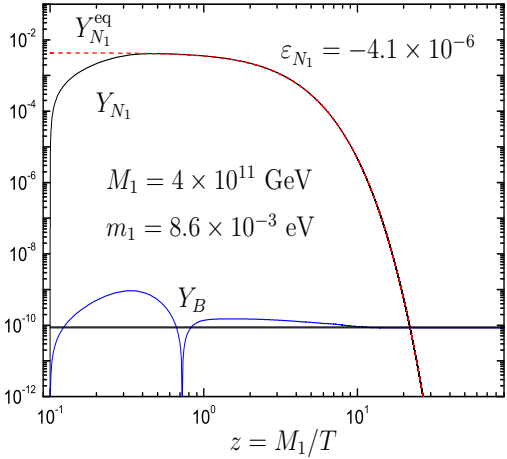, bbllx=6.5cm, bblly=6.0cm, bburx=15.5cm, bbury=16.0cm,%
width=7cm, height=9cm, angle=0, clip=0} \vspace{-0.5cm}
\caption{\label{yb} The solutions of the Boltzmann equations.
$Y^{(\textrm{eq})}_{N_{1}}$ is the (equilibrium) number of heavy
Majorana neutrino, while $Y^{}_{B}=28/79\,Y^{}_{B-L}$ denotes the
baryon. In the numerical calculations, we have taken
$M_{1}^{}=4\times 10^{11}\,\textrm{GeV}$, $m^{}_1 = 8.6 \times
10^{-3} ~ {\rm eV}$, $\Delta m^2_{21} = 8.3 \times 10^{-5} ~ {\rm
eV}^2$, $\Delta m^2_{32} = 2.3 \times 10^{-3} ~ {\rm eV}^2$,
$y_0=0.02$, $r=0.64$ and $(\beta^{}_\nu, \gamma^{}_\nu) =
(-100^\circ, -36^\circ)$. The predicted value for the final baryon
asymmetry is $Y_{B}\simeq 8.5\times 10^{-11}$.}
\end{figure}

We now calculate the produced baryon asymmetry in the minimal type
II seesaw model. Assuming the heavy Majorana neutrino $N_{1}$ is
much lighter than the triplet Higgs $\Delta_{L}$, the final lepton
asymmetry, which is partially converted to the baryon asymmetry via
the sphaleron process \cite{krs1985}, will mainly come from the
decay of $N_{1}$. We can write down the relevant Boltzmann
equations:
\begin{eqnarray}
\frac{dY_{N_{1}}^{}}{dz}&=& -\frac{z}{sH(M_{1})}\left[\gamma_{D}^{}+2\gamma_{\phi,s}^{}+4\gamma_{\phi,t}^{}\right]\nonumber\\
&&\times \left(\frac{Y_{N_{1}}^{}}{Y_{N_{1}}^{\rm eq}}-1\right)\,,\\
\frac{dY_{B-L}^{}}{dz}&=&-\frac{z}{sH(M_{1})}\left[\varepsilon_{N_{1}}^{}\gamma_{D}^{}\left(\frac{Y_{N_{1}}^{}}{Y_{N_{1}}^{\rm eq}}-1\right)\right.\nonumber\\
&&+\left.\frac{Y_{B-L}^{}}{Y_{l}^{\rm eq}} \times
\left(\frac{1}{2}\gamma_{D}^{}+2\gamma_{N}^{}+2\gamma_{N,t}^{}\right.\right.\nonumber\\
&&+\left.\left.
2\gamma_{\phi,t}^{}+\frac{Y_{N_{1}}^{}}{Y_{N_{1}}^{\rm
eq}}\gamma_{\phi,s}^{}\right)\right]\,,
\end{eqnarray}
where $z=M_{1}/T$, $H(M_{1})$ is the Hubble parameter at $T=M_{1}$,
$Y_{X}^{(\rm eq)}$ is defined as $Y_{X}^{(\rm eq)}=n_{X}^{(\rm
eq)}/s$ with $n_{X}^{(\rm eq)}$, $s$ being the (equilibrium) number
density and the entropy density, respectively. $\gamma_{X}$, which
has been calculated by many authors \cite{GU}, is the reaction
density of the relevant process.

As pointed out in the foregoing discussions, because there are no
other heavy Majorana neutrinos $N_{i}\, (i\neq 1)$, the CP
asymmetry $\varepsilon_{N_{1}}$ just comes from the interference
between the tree level diagram and the vertex correction induced
by the triplet Higgs $\Delta_{L}$. For $M^{}_{\Delta} \gg
M^{}_{1}$, we obtain:
\begin{eqnarray}
\varepsilon_{N_{1}}^{} & \equiv &  \varepsilon_{N_{1}}^{\Delta}
\simeq\frac{3}{16 \pi} \frac{M_{1}}{v^{2}}
\frac{\sum_{\alpha\beta} \text{Im}\left[y^{\nu
\dagger}_{1\alpha}y^{\nu \dagger}_{1\beta}
\left(M^{\textrm{II}*}_{\nu}\right)_{\alpha\beta}\right]}{\left(y^{\nu
\dagger}y^{\nu}\right)_{11}}\nonumber \\
&=& \frac{3}{16 \pi} \frac{y^2_0}{1+r^2} \left [ \widehat{A} \sin
\gamma^{}_\nu + 2\widehat{B} r^2 \sin\beta_\nu\right ] \,.
\end{eqnarray}
In the last step, Eqs. (\ref{massII}) and (\ref{yukawa}) have been
used.

Given appropriate parameters consistent with the neutrino
oscillations, both the CP asymmetry $\varepsilon_{N_{1}}$ and the
reaction density $\gamma_{X}$ can be computed, and then the
Boltzmann equations can also be solved. For example, we take
$M_{1}^{}=4\times 10^{11}\,\textrm{GeV}$, $\Delta m^{2}_{21} = 8.3
\times 10^{-5} ~ {\rm eV}^{2}$, $\Delta m^{2}_{32} = 2.3 \times
10^{-3} ~ {\rm eV}^{2}$, $m^{}_{1} = 8.6 \times 10^{-3} ~ {\rm
eV}$, $y_0=0.02$, $r=0.64$, $(\beta^{}_\nu, \gamma^{}_\nu) =
(-100^\circ, -36^\circ)$ and find the current baryon asymmetry in
the universe can be successfully explained. The numerical results
are shown in Fig. \ref{yb}. Here the relation
$Y^{}_{B}=28/79\,Y^{}_{B-L}$ \cite{ks1988} has been adopted.
During the calculations, the parameters $\widehat{A}$,
$\widehat{B}$ and $\widehat{C}$, which characterize the
contribution of the triplet Higgs to the effective neutrino mass
matrix (\ref {massNeutrino}), can also be determined by using Eq.
(\ref{ABCnu}): $\widehat{A}\simeq 0.41$, $\widehat{B}\simeq
-0.093$ and $\widehat{C}\simeq 0.38$. In this case, it is easy to
see that the right-handed neutrino provides a sizable contribution
to the light neutrino masses.

In summary, we have proposed a minimal type-II seesaw model, which
contains only one right-handed neutrino in addition to the triplet
Higgs. Different from the usual type II seesaw models with several
right-handed neutrinos, our model gives the CP asymmetry in the
decay of the lightest right-handed neutrino just by the vertex
correction involving the triplet Higgs, no matter whether the
contribution of the triplet Higgs dominates the light neutrino
masses. For illustration, the Fritzsch-type lepton mass matrices are
considered. The neutrino oscillations and the baryon asymmetry can
simultaneously be well explained. More phenomenological studies in
this scenario are apparently interesting and desirable.

\begin{acknowledgments}
The authors would like to thank Wan-Lei Guo for useful
discussions. This work is supported in part by the Ministry of
Science and Technology of China under Grant No. NKBRSF G19990754
and by the National Natural Science Foundation of China. Gu would
like to thank the hospitality of Department of Engineering
Physics, Tsinghua University where part of this work was finished.
\end{acknowledgments}

\end{document}